\documentclass[%
 reprint,
 showpacs,
 amsmath,amssymb,
 aps,
 prl,
]{revtex4-1}

\usepackage{graphicx}
\usepackage{dcolumn}
\usepackage{bm}
\usepackage{epstopdf} %
\usepackage{xcolor}

\begin{document}

\preprint{APS/123-QED}

\title{Evidence for GeV Pair Halos around Low Redshift Blazars\\}%

\author{Wenlei Chen}
 \email{wenleichen@wustl.edu}
\author{James H. Buckley}%
\author{Francesc Ferrer}%
\affiliation{%
 Department of Physics and McDonnell Center for the Space Sciences, Washington University in St. Louis, MO 63130, USA.\\
}%

\date{\today}

\begin{abstract}
We report on the results of a search for $\gamma$-ray pair halos with a stacking analysis of low-redshift blazars using data from the Fermi Large Area Telescope. For this analysis we used a number of \emph{a-priori} selection criteria, including the spatial and spectral properties of the Fermi sources. The angular distribution of $\sim$ 1GeV photons around 24 stacked isolated high-synchrotron-peaked BL Lacs with redshift $z<0.5$ shows an excess over that of point-like sources. A statistical analysis yields a Bayes factor of $\mathrm{log}_{10}B_{10}>2$, providing evidence in favor of extended emission against the point-source hypothesis, consistent with expectations for pair halos produced in the IGMF with strength $B_{\mathrm{IGMF}}\sim 10^{-17}-10^{-15}\mathrm{G}$.
\end{abstract}

\maketitle

\section{\label{sec:intro}Introduction}

The magnetic fields that are observed in galaxies and galaxy clusters are believed to result from the dynamo amplification of weak magnetic field seeds, whose origin remains a mystery. Intergalactic magnetic fields (IGMFs), deep in the voids between galaxies, provide the most accurate image of the weak primordial seed fields and could be linked to the early stages in the evolution of the universe (see e.g. \cite{Kandus2011} for a recent review). Among the several methods used to study cosmological magnetic fields (see e.g.\cite{Durrer2013} for a recent review), the observation (or nondetection) of cascade emission from blazars can potentially measure very weak IGMFs. A number of blazars have been observed to emit both very-high-energy (VHE, $>100$ GeV) $\gamma$-rays with ground-based $\gamma$-ray instruments and high-energy (HE, MeV/GeV) $\gamma$-rays with the Fermi Gamma-ray Space Telescope \cite{Nolan2012,Ackermann2013a}. Most of the detected TeV $\gamma$-rays are from the nearest sources since such high energy $\gamma$-rays cannot propagate over long distances in intergalactic space due to interactions with the extragalactic background light (EBL). Of course, some higher-redshift sources still have detectable TeV emission (e.g. blazar PKS1424+240, which has redshift lower limit of $z>0.6$ \cite{Furniss2013}), but with highly absorbed spectra consistent with theoretical calculations of the attenuation by the EBL. \cite{Aharonian2013,Essey2010a,Essey2010b,Essey2011a,Essey2012}. These interactions of TeV $\gamma$-rays with the EBL produce electron-positron pairs that subsequently are cooled by inverse Compton (IC) interactions with the Cosmic Microwave Background (CMB), ultimately leading to GeV $\gamma$-ray emission from these pair cascades. Since magnetic fields deflect the electron-positron pairs changing the angular distribution of cascade emission, searches for extended GeV emission around blazars can provide an avenue for constraining the IGMF.

Due to the low GeV $\gamma$-ray flux from extragalactic sources, it is difficult to examine the angular extent of the photon events from a single blazar or even to assess the joint likelihood for detailed fits to a set of individual sources where individual source parameters are taken to be completely independent. To overcome this limitation, stacking sources has been used to make such statistical analysis feasible. Despite early hints at a signal in the stacking analysis of 170 brightest active galactic nuclei (AGNs) using 11-month Fermi observations \cite{Ando2010}, by comparing with the GeV emission from the Crab Nebula [which is essentially a point source for the Fermi Large Area Telescope (Fermi-LAT)], A. Neronov et al. \cite{Neronov2011} found no significant evidence of extended emission and argued that the apparent excess could be attributed to an underestimation of the real PSF \cite{Abdo2009}. A subsequent analysis by Ackermann et al. \cite{Ackermann2013b} comparing an updated PSF to one hundred stacked BL Lac AGNs did not find any statistically significant halo-emission either.

The cascade emission from individual blazars has also been studied by modeling the intrinsic TeV spectra and adopting EBL and cosmological microwave background (CMB) models (e.g. \cite{Murase2008,Neronov2010,Essey2011b,Arlen2012,Tanaka2014}). Delays in arrival time of the cascade emission were used to explain the non-dectection of several TeV sources in Fermi energy, and to derive a lower bound of the IGMF strength (e.g. $\sim10^{-20}-10^{-19}\mathrm{G}$ in \cite{Murase2008}). The angular extent of the cascade signals caused by IGMFs above
$\sim10^{-16}\mathrm{G}$ also provided an explanation for the non-detection of TeV sources 1ES 0229+200 and 1ES0347-121 by Fermi \cite{Neronov2010}. W. Essey et al. reported a possible \emph{measurement} of IGMFs in the range $1\times10^{-17}-3\times10^{-14}\mathrm{G}$\cite{Essey2011b} based on the TeV-GeV spectra. Very recently, a study of 1ES0347-121 spectral energy distribution (SED) provided an IGMF estimation of $3\times10^{-17}\mathrm{G}$ \cite{Tanaka2014}. Fitting to TeV data from, e.g., VERITAS and HESS, such studies yielded detailed predictions of the cascade emission, but invariably made assumptions about the sources, e.g. the relationship of the long-term TeV emission to measurement of a few flares. The upper bound of the IGMF strength with correlation length above $\sim 1$ Mpc is below $\sim 10^{-9}\mathrm{G}$ constrained by the non-detection of the large scale CMB anisotropies, and is given to be $\sim 10^{-12}\mathrm{G}$ by the galaxy cluster simulation, as summarized in \cite{Neronov2009,Neronov2010}. The likely range of the IGMF strength from previous studies is given from $\sim 10^{-20}\mathrm{G}$ to $\sim 10^{-12}\mathrm{G}$.

As the energies of the primary $\gamma$-rays increase, the pair production occurs closer to the source, reducing the angular size of the cascade. Depending on the strength of the IGMF and the redshift of the source, the highest energy emission might not be resolved by the Fermi PSF. While at lower energies (especially for the nearest sources), the emission may be too diffuse to be readily detected. It follows that only a few blazars would have cascade emission that can be statistically detected through their angular profiles.

In our study, we combine data from 24 isolated high-synchrotron-peaked (HSP) BL Lacs which are \emph{a-priori} selected to provide the best prospects for detection and adequate photon counting statistics. Both frequentist likelihood ratio test (LRT) statistics and Bayes factors are evaluated for estimating the pair halo parameters (the angular size and halo fraction), which consequently provide the possible range of IGMF strength.

\section{Data Preparation and Selection Criteria for Stacking Sources}

We use the Fermi-LAT Pass 7 reprocessed data through February 2014: SOURCE class front-converted photon events are binned into four logarithmically spaced energy ranges to roughly equalize counts (see Table \ref{Tab:1}). The source candidates are selected from the AGN associated sources in the Fermi-LAT High-Energy Catalog (1FHL \cite{Ackermann2013a}). The regions of the Galactic disk and Fermi bubbles are excluded to avoid anisotropic background emission \cite{supp}.

Data is also divided into angular bins to provide adequate statistics. Source bins of equal solid angle are set around the direction of the source, surrounded by a larger background bin with an outer boundary of $5^\circ$. To reduce systematic errors from nearby sources, we require that no nearby sources (those bright enough to appear in the 2FGL catalog) are within $2.3^\circ$ of the stacked sources and correct for the impact of any remaining nearby sources by defining an exclusion region of radius $\theta_{cut}$ ($=2.3^\circ$) about these sources; we account for these exclusion regions by assuming that the signal and background effective area is reduced in proportion to the excluded solid angle. The size of the source bins $\theta_{in}$, is a function of energy chosen to be greater than the 95\% containment angle of the PSF in the corresponding energy range \cite{IRF} (see Table \ref{Tab:1}).

\begin{table}[htbp]
	\centering
	\caption{Energy bins and values of $\theta_{in}$}
	\begin{tabular}{ ccccc }
		\hline
		Energy (GeV) & $~~1-1.58$ & $~1.58-3.16$ & $~3.16-10$ & $~10-100$ \\
		$\theta_{in}$ & $2.3^\circ$ & $1.6^\circ$ & $1^\circ$ & $0.8^\circ$ \\
		\hline
	\end{tabular}
	\label{Tab:1}     
\end{table}

Assuming that the correlation length of the IGMF is much greater than the mean free path for IC scattering ($\sim10^1-10^2$ kpc, see detailed discussion in \cite{Neronov2009}, also in \cite{Tashiro2013}), we estimate the typical size of a pair halo to be
\begin{equation}
\begin{aligned}
\Theta(E_\gamma,z_s,B_0)\approx 9.2\times 10^{-4}\left[1+z_{\gamma\gamma}(E_\gamma,z_s)\right]^{-2}\\
\times\left(\frac{E_\gamma}{100\mathrm{GeV}}\right)^{-1}\left(\frac{B_0}{10^{-16}\mathrm{G}}\right)\left[\frac{d_\gamma(E_\gamma,z_s)}{d_s(z_s)}\right],
\end{aligned}
\label{Eq:1}
\end{equation}
where $E_\gamma$ is the energy of the cascade photon observed by Fermi, $z_s$ is the observed redshift of the source, and $B_0$ is the field strength at the present epoch. To get the estimate above, we followed the discussion in A. Neronov and D. V. Semikoz \cite{Neronov2009} (see also \cite{Tashiro2013}), where $z_{\gamma\gamma}$ is the redshift of pair production, $d_\gamma$ and $d_s$ are the commoving mean free path for pair production and the commoving distance to the source, respectively \cite{supp}. Given the finite Fermi PSF, it is quite unlikely to detect the extended emission from high-redshift sources. For example, from Eq. \ref{Eq:1}, an IGMF of $\sim 10^{-16}\mathrm{G}$ would result in a halo of angular radius of $\sim 2^\circ$ at $1\mathrm{GeV}$ for a source at $z=0.3$. If the same source were located at $z=0.8$, the halo size would decrease to $\sim 0.2^\circ$, which is much smaller than the Fermi PSF and would appear like a point source. In addition, most of the sources from $z<0.5$ would be seen along the lines of sight that do not cross astrophysical systems (i.e. galaxy and galaxy clusters) which host large magnetic fields \cite{Aharonian2013}, indicating that the cascade emission from these sources is most likely produced in the intergalactic space.

Both observational and theoretical arguments lead us to expect that HSP BL Lac objects are the most likely sources of the VHE $\gamma$-rays needed to produce the GeV cascades. For example, in \cite{Ghisellini1998,Fossati1998}, we see a strong correlation of the occurrence of a HSP energy with TeV emission. This is naturally explained if the same population of VHE electrons that produce the X-ray synchrotron radiation also produce the TeV $\gamma$-rays by IC in the source region (e.g. AGN jets). For this study, we subdivide data into Flat Spectrum Radio Quasars (FSRQs), and BL Lac objects. Since the FSRQs are typically very distant sources with lower-energy synchrotron peaks (LSP), we expect these sources to lack observable GeV pair halos, serving as a control population.

\section{\label{sec:map}Distribution of the GeV $\gamma$-rays around Stacked Blazars}

\begin{figure}[htbp!]
	\centering
	\includegraphics[width=8cm, angle=0]{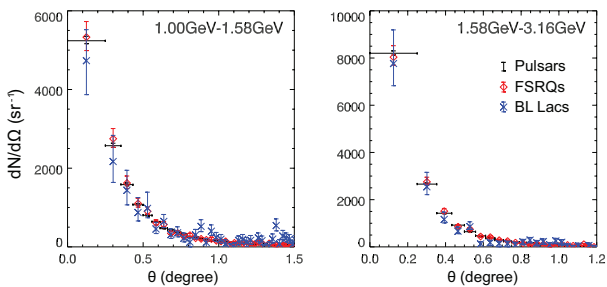}
	\caption{\label{Fig:1}Angular distribution of photon events around the stacked pulsars (black), the stacked FSRQs (red), and the stacked BL lacs (blue): vertical errors are the 68\% confidence intervals; horizontal errors show the size of angular bins.}
\end{figure}

We identify 24 HSP BL Lacs with redshift $z<0.5$ that satisfy our selection criteria and stack their photon events. As a control population, 26 FSRQs (with any redshift) are also selected by the same criteria. As evident in past searches for pair halos, a thorough understanding of the PSF is critical for this type of study. Pulsars with unresolved pulsar wind nebulae (PWN) can be used as calibration sources since they are effective point sources for Fermi-LAT \cite{Neronov2011,Ackermann2013b}; here we choose the Geminga \cite{Ackermann2011} and Crab \cite{Hester2008} pulsars. To plot different angular distribution profiles of different stacked source classes, we calculate and remove the diffuse background for each source, sum the background-subtracted counts and then normalize the profiles. We calculate the angular profiles for the stacked pulsars, the 24 BL Lacs, and the 26 FSRQs, as shown in Fig. \ref{Fig:1}. The angular profiles for stacked pulsars agree with their PSFs (P7REP\_SOURCE\_V15) in each energy range \cite{supp}. The normalized angular profiles of stacked BL Lacs have lower scaled counts per unit solid angle at small $\theta$, providing evidence for extended emission since the additional counts in the extended halo reduce the scaled counts at small angles after normalization. The deficit in counts at small $\theta$ (evidence for extended emission) is only significant in the lowest energy bin, consistent with the expectation that the angular extent of the halo is larger at lower energies, as indicated in Eq. \ref{Eq:1}. In contrast, the angular profiles of the stacked FSRQs are indistinguishable from our surrogate point-source data from pulsars, as shown in Fig. \ref{Fig:1}.

\section{\label{sec:stat}Statistical Evidence for Pair-Halo Emission and Estimation of the IGMF}

To model the normalized angular profiles $g(\theta)$, we use
\begin{equation}
g(\theta;f_{\mathrm{halo}},\Theta)=f_{\mathrm{halo}}g_{\mathrm{halo}}(\theta;\Theta)+(1-f_{\mathrm{halo}})g_{\mathrm{psf}}(\theta),
\label{Eq:2}
\end{equation}
where $f_{\mathrm{halo}}$ is the fraction of the pair halo component, $\Theta$ is a single parameter characterizing the angular extent of the halo. $g_{\mathrm{psf}}(\theta)$ is the effective PSF for the stacked source \cite{supp} and $g_{\mathrm{halo}}(\theta;\Theta)$ is a Gaussian function of $\theta$ (in the small angle approximation) convolved with the PSF. Then, the number of photon events in the $j$-th angular bin around the stacked source is estimated by
\begin{equation}
\lambda_j(f_{\mathrm{halo}},\Theta,\bm{\mu},\bm{A})=\sum_{i}(A_ig_j+\mu_i)\Omega_{i,j}w_{i,j},
\label{Eq:3}
\end{equation}
where $g_j$ is the discrete value of the normalized angular distribution $g(\theta)$ given by Eq. \ref{Eq:2}, $\bm{A}$ and $\bm{\mu}$ are a set of normalization factors $\{A_i\}$ and a set of the assumed uniform background values (in counts per unit solid angle) $\{\mu_i\}$, respectively, for each of the $i$-th source. $\Omega_{i,j}$ is the solid angle of the $j$-th angular bin around the $i$-th source. $w_{i,j}=\mathcal{E}_{i,1}/\mathcal{E}_{i,j}$ is the exposure corrector to calibrate the expected counts in the $j$-th angular bin around the $i$-th source to the level of the center angular bin of this source, where $\mathcal{E}$ is the averaged exposure of the angular bin. For a given configuration of the angular bins, a set of estimators $\{\lambda_j\}$ is a function of $f_{\mathrm{halo}}$, $\Theta$, $\bm{\mu}$, and $\bm{A}$.

We present both a frequentist and Bayesian analysis of the data. A set of observed counts $\bm{N}=\{N_{i,j}\}$ are estimated by the model given by Eq. \ref{Eq:3}, where $N_{i,j}$ is the number of counts in the $j$-th angular bin around the $i$-th source. Counts in the background bins are also estimated by the isotropic background model derived from $\bm{\mu}$. For the frequentist analysis, maximum likelihood estimation (MLE) is used for the model fitting. The logarithm of the likelihood ratio is evaluated as a test statistic (TS), providing the confidence level of getting $\bm{N}$. Since the number of counts in each bin ($i$, $j$) can be quite small (average counts in an individual bin $\left<N_{i,j}\right>\sim3$ for the BL Lacs and $\left<N_{i,j}\right>\sim6$ for the FSRQs in the 1 GeV-1.58 GeV energy range), a naive application of the MLE where one evaluates the joint likelihood $\mathcal{L}\equiv\prod_{i,j}P(N_{i,j}|\lambda_{i,j})$ can give large estimation errors, resulting in a non-converging distribution of the TS (the logarithmic likelihood ratio), and can potentially lead to a type II error \cite{supp}. While this is addressed by the Bayesian analysis, it may be a problematic for a frequentist inference \cite{Baldwin2013}. Here we adopt a novel approach \cite{supp} where we repartition the data into two sets: the stacked angular distribution $\{\sum_{i=1}^{n}N_{i,j}\}\equiv\{\eta_j\}$ obtained by summing over sources $i$, and the stacked source distribution $\{\sum_{j=1}^{m}N_{i,j}\}\equiv\{\zeta_i\}$ obtained by summing over angular bins $j$, where $m$ and $n$ are the total number of angular bins and stacked sources, respectively. The likelihood of obtaining $\{\zeta_i\}$ and $\{\eta_j\}$ is calculated as $\mathcal{L}_{\rm{on}}$. This is combined with the likelihood of getting a set of $\{N_{i,m}\}$ counts detected in each background bin around each source $\mathcal{L}_{\rm{off}}$.

\begin{figure}[htbp!]
	\centering
	\includegraphics[width=8cm, angle=0]{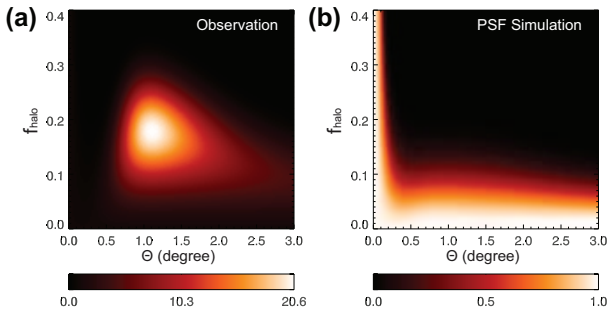}
	\caption{\label{Fig:2}Likelihood ratio maps in the 1GeV-1.58GeV energy bin. Colors show the ratio of the likelihood of the extended-emission hypothesis to that of the null hypothesis (the PSF). (a) Likelihood ratio maps for stacked BL Lacs; (b) Likelihood ratio maps for a point source with angular distribution given by the PSF and with total number of events in each energy bin set equal to that of the stacked BL Lacs.}
\end{figure}

We subsequently evaluate the joint likelihood $\mathcal{L}=\mathcal{L}_{\rm{on}}\times\mathcal{L}_{\rm{off}}$ which is defined in the multidimensional space of the model parameters, $\bm{x}=(f_{\mathrm{halo}},\Theta,\bm{\mu},\bm{A})$ \cite{supp}. Note that both $\zeta_i$ and $\eta_j$ have relatively large number of counts, and $N_{i,m}$ is also relatively large since the solid angle of the background bins is much larger than that of an individual angular bin ($i$, $j$), hence the following frequentist analysis acting on $\zeta_i$, $\eta_j$, and $N_{i,m}$ will not encounter the problem of small sample size. To get the quantitative significance of the pair halo, we focus on the space of the two model parameters, $f_{\mathrm{halo}}$ and $\Theta$. We must distinguish between two hypotheses in this space: the hypothesis of halo emission $\mathcal{H}_1$ and the null hypothesis $\mathcal{H}_0$, where $\mathcal{H}_0$ denotes a pure point source where either $f_{\mathrm{halo}}=0$ or $\Theta=0$, and for $\mathcal{H}_1$, the two parameters are free. The ratio of the maximum likelihood of $\mathcal{H}_1$ for a given pair of $f_{\mathrm{halo}}$ and $\Theta$ to that of $\mathcal{H}_0$ is evaluated and displayed in ($f_{\mathrm{halo}},\Theta$)-space. Fig. \ref{Fig:2} shows the likelihood ratio maps for the stacked BL Lacs (a) and the maps for the simulated point source (labeled PSF) with total number of events in each energy bin set to that of the stacked BL Lacs (b). From Eq. \ref{Eq:2}, $\mathcal{H}_0$ gives $g(\theta)=g_{\mathrm{psf}}(\theta)$, indicating that any point on $f_{\mathrm{halo}}$ and $\Theta$ axes in each map gives a constant likelihood corresponding to a null model without extended emission. Fig. \ref{Fig:2}(b) shows that the maximum values of the likelihood ratio are distributed along the axes, consistent with the null hypothesis.

The 1 GeV-1.58 GeV likelihood ratio map shows a peak at non-zero $f_{\mathrm{halo}}$ and $\Theta$ (Fig. \ref{Fig:2}). In the higher energy bins \cite{supp}, the highest likelihood appears close to the $f_{\mathrm{halo}}$ and $\Theta$ axes (where the null model is located). The fact that the likelihood maps for the higher energy bins are consistent with the null hypothesis matches our expectation based on the angular distribution measurements shown in Fig. \ref{Fig:1}, where no significant difference is seen between the profiles of stacked pulsars and stacked BL Lacs in the plots of the higher energy bins. From the distributions of the maximum values of the likelihood ratio, the pulsars are shown to appear as point sources for Fermi-LAT \cite{supp}. To put our results in the more familiar language of frequentist statistics, we simulated the distribution of the TS by using a Monte Carlo method based on the null hypothesis. The LRT shows that if the stacked source appears to be a point source given by the Fermi PSF, the significance (probability) of getting an observation of the stacked BL Lacs in the 1GeV-1.58GeV energy bin is equivalent to the significance (probability) of getting a normal distributed sample at $\sim 2.3\sigma$ \cite{supp}.

Alternatively, we calculate the Bayes factors $B_{10}=\mathcal{L}_B(H_1|\bm{N})/\mathcal{L}_B(\mathcal{H}_0|\bm{N})$ \cite{Kass1995,Protassov2002} to test the extended-emission hypothesis $H_1$ for given values of $f_{\mathrm{halo}}=f^*_{\mathrm{halo}}$ and $\Theta=\Theta^*$ (a subset of $\mathcal{H}_1$) against the null hypothesis $\mathcal{H}_0$ \cite{supp}. For hypotheses $\bm{H}=\{\mathcal{H}_0,H_1\}$, the Bayesian likelihood $\mathcal{L}_B$ is given by
\begin{equation}
\mathcal{L}_B(\bm{H}|\bm{N})=\int\mathrm{d}\bm{x}P(\bm{N}|\bm{x},\bm{H})\pi(\bm{x}|\bm{H}).
\label{Eq:4}
\end{equation}
Different from the frequentist LRT, for a Bayesian method, the problem of limited statistics in the ($i$, $j$) bins is eliminated \cite{Baldwin2013}, and we can include all the information contained in the data. We are left with the straight forward (but computationally difficult) task of evaluating the multi-dimensional integral over model parameters to obtain the p-value. In Eq. \ref{Eq:4}, the prior can be designed to constrain the total number of counts with no additional assumptions, while the posterior density is given by the joint Poisson likelihood of getting the observation $\bm{N}$ \cite{supp}. 

\begin{figure}[htbp!]
	\centering
	\includegraphics[width=8cm, angle=0]{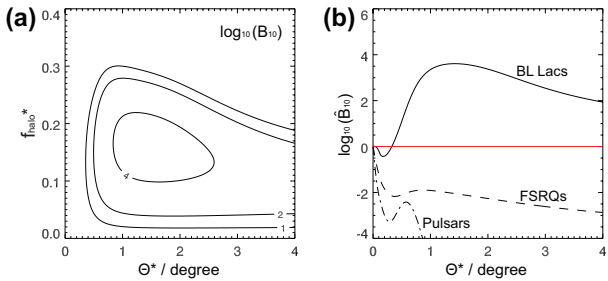}
	\caption{\label{Fig:3}Bayes factors in the 1GeV-1.58GeV energy bin. (a) Bayes factors of the hypotheses $H_1 (f_{\mathrm{halo}}=f^*_{\mathrm{halo}},\Theta=\Theta^*)$ against $\mathcal{H}_0 (\{f_{\mathrm{halo}}=0\}\bigcup\{\Theta=0\})$ for the stacked BL Lacs; (b) Bayes factors of the hypotheses $\hat{H}_1 (f_{\mathrm{halo}}\in(0,1],\Theta=\Theta^*)$ against $\mathcal{H}_0$ for the stacked BL Lacs (solid line), FSRQs (dashed line), and pulsars (dash-dot line).}
\end{figure}

We plot the contours of $\mathrm{log}_{10}B_{10}$ in the $f^*_{\mathrm{halo}}$-$\Theta^*$ coordinates, as shown in Fig. \ref{Fig:3} (a). We find that $\mathrm{log}_{10}(B_{10})>2$ for the 1GeV-1.58GeV energy bin, showing decisive evidence \cite{Kass1995} for the hypothesis of extended emission against the null hypothesis. While $\mathrm{log}_{10}(B_{10})<0.5$ at higher energies, providing no significant evidence against the null hypothesis. The information about the IGMF is contained in the extended emission. Here we focus on the model factor $\Theta$, and seek to get the quantitative significant range of its values for the stacked BL Lacs. We introduce a hypothesis $\hat{H}_1$ for a given $\Theta^*$ with all possible values of $f_{\mathrm{halo}}$. The Bayes factors of $\hat{H}_1$ can be evaluated by integrating the Bayesian likelihood $\mathcal{L}_B$ over all possible values of $f_{\mathrm{halo}}$ \cite{supp}. Thus, the resulting Bayes factors $\hat{B}_{10}$ of $\hat{H}_1$ against $\mathcal{H}_0$ are given as a function of $\Theta^*$, as shown in Fig. \ref{Fig:3} (b). From the Bayes factors, we obtain the values of $\Theta$ given by the most likely hypothesis (where $\mathrm{log}_{10}(B_{10})>2$): $\sim 0.6^\circ-4^\circ$ in the first energy bin. Recalling Eq. \ref{Eq:1}, using the average redshift of the stacked BL Lacs $\left\langle z\right\rangle\approx0.23$, the strength of IGMF is conservatively estimated to be in the range of $B_{\mathrm{IGMF}}\sim 10^{-17}-10^{-15}\mathrm{G}$. These values are larger than the lower bound derived from observations of 1ES0347-121 in \cite{Neronov2010} and consistent with the results in \cite{Ando2010,Essey2011b,Tanaka2014}. The negative Bayes factors for the stacked pulsars and FSRQs [as shown in Fig. \ref{Fig:3} (b)] provide no evidence for pair halos, consistent with the results given by the frequentist LRT.  

\section{\label{sec:discussion}Discussion}

In this study, we presented an analysis of the angular distribution of $\gamma$-rays from a subset of sources selected \emph{a-priori} to minimize systematics but maximize chances of finding spatially resolved halo emission. This study provides an interesting hint of a detection of pair halos, shown both by a frequentist and a Bayesian analysis, resulting in a possible measurement of the IGMF, consistent with prior limits.
	
Most of the Fermi sources have nearby sources (within $2^\circ$), which will contaminate the stacked angular profiles. Previous studies restricted the energy range to be greater than 1 GeV to limit the contamination. However, this criterion is only valid in stacking the brightest sources and analyzing their angular photon-distribution. While HSP BL lacs are the most likely halo sources, they are not the brightest sources for Fermi-LAT. Moreover, the containment angle of the PSF at 1 GeV is $\sim 1^\circ$, large enough to still allow contamination from nearby sources for many of these AGNs.

For the entire map, the modeled source counts leaking into the background region fall below one standard deviation of the background counts $\sim \sqrt{N_{bg}}$. Hence we are justified in neglecting spillover into the background region. We choose not to use the Fermi diffuse background models in this study, because the empirical model could contain contributions from extended sources and other assumptions about the angular distribution of the emission from local sources. We selected blazars in the Galactic polar regions, where we assumed that the diffuse background observed around our selected sources is isotropic (see \cite{supp}. To test the sensitivity of our assumption of a uniform background, we compared the isotropic background model with the Fermi diffuse background model, showing no significant evidence in favor of the Fermi diffuse model).

Given the limitations of the stacking-source method, only an average range of the IGMFs can be recovered. In a finite sky-region, the emission from very large halos will be taken into account in our statistical analysis as background counts, because our method is insensitive to very large pair halos, whose photon fluxes are too extended to be resolved from the background emission. Since the maximum angular search window is limited by source confusion and other experimental factors, we can not provide as strong a constraint on the maximum allowed angular extent of the GeV $\gamma$-ray emission and the maximum field strength as we can on the minimum angular extent and field strength (as shown in Fig. \ref{Fig:3}, where a long tail of significance can be seen at large angles). In addition, the small-angle approximation implicit in Eq. \ref{Eq:1} might not hold for the larger magnetic fields, since the electron-positron pairs might follow trajectories with complete loops \cite{Long2015}. Thus, the estimation of IGMFs in this study is still marginally consistent with the results from Tashiro et al. \cite{Tashiro2014}, in which the strength of the helical component of the IGMF is given as $\sim 10^{-14}\mathrm{G}$ by analyzing the Fermi extragalactic diffuse background.

\begin{acknowledgments}
Authors acknowledge the Fermi team for providing the Fermi-LAT data (available in the Fermi Science Support Center, http://fermi.gsfc.nasa.gov/ssc/). W. Chen thanks the Department of Physics, Washington University in St. Louis (WUSTL) for awarding the Arthur L. Hughes Fellowships to support his study.  J. H. Buckley and F. Ferrer have been partially supported by the U. S. Department of Energy grant DE-FG02-91ER40628 at WUSTL.	
\end{acknowledgments}


\clearpage

\appendix
\renewcommand{\theequation}{S-\arabic{equation}}
\setcounter{equation}{0}  

\widetext

\begin{center}
	{\bf \large Evidence for GeV Pair Halos around Low Redshift Blazars}
	\vspace*{0.1in}
	
	Wenlei Chen, James H. Buckley, Francesc Ferrer\\
	\vspace*{0.05in}
	\emph{\small Department of Physics and McDonnell Center for the Space Sciences,
		Washington University in St. Louis, MO 63130, USA.}\\
	\vspace*{0.1in}
	\bf \large Supplemental Material
\end{center}

\section{\sc \large Details of Data Selection}

\begin{figure}[htbp]
	\centering
	\includegraphics[width=16cm, angle=0]{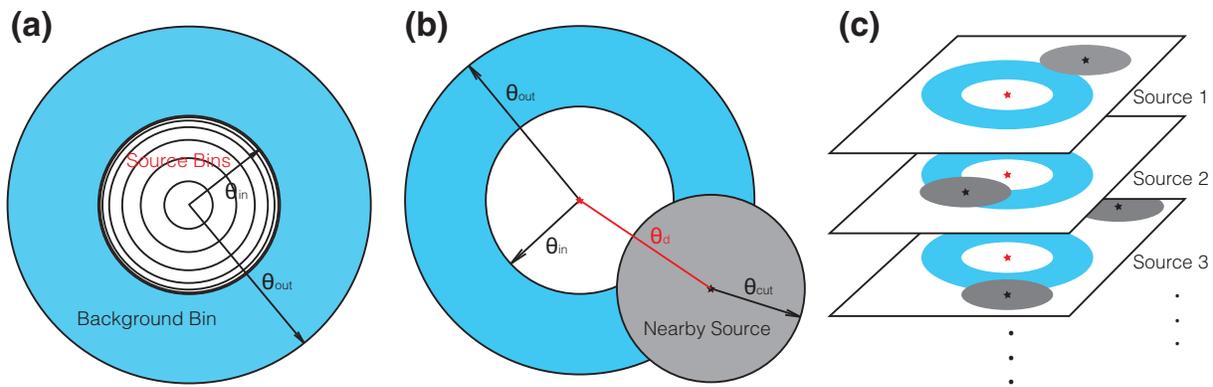}
	\caption{\small Configuration of angular bins: (a) Angular source bins with the same angular size (solid angle) and the background bin. (b) Definition of parameters $\theta_d$ and $\theta_{cut}$ used in excluding sources with very nearby sources. (c) Masking out the nearby sources in stacking the blazars.}
	\label{Fig:S1}
\end{figure}

The angular distribution of the GeV emission is measured by counting the number of photon events in angular bins chosen as depicted in Figure \ref{Fig:S1}(a): source bins of equal solid angle are set around the direction of the source, surrounded by a larger background bin. Before measuring the angular distribution of the $\gamma$-ray emission around each source, the following assumptions are made: 1) The photon events are distributed with azimuthal symmetry with respect to the direction of the source; 2) The $\gamma$-ray flux in the background bin is assumed to be dominated by a diffuse background and can therefore be used to estimate the background in the source bins; 3) The background $\gamma$-ray flux is uniformly distributed in the detection region within all the angular bins.

\begin{figure}[htbp!]
	\centering
	\includegraphics[width=16cm, angle=0]{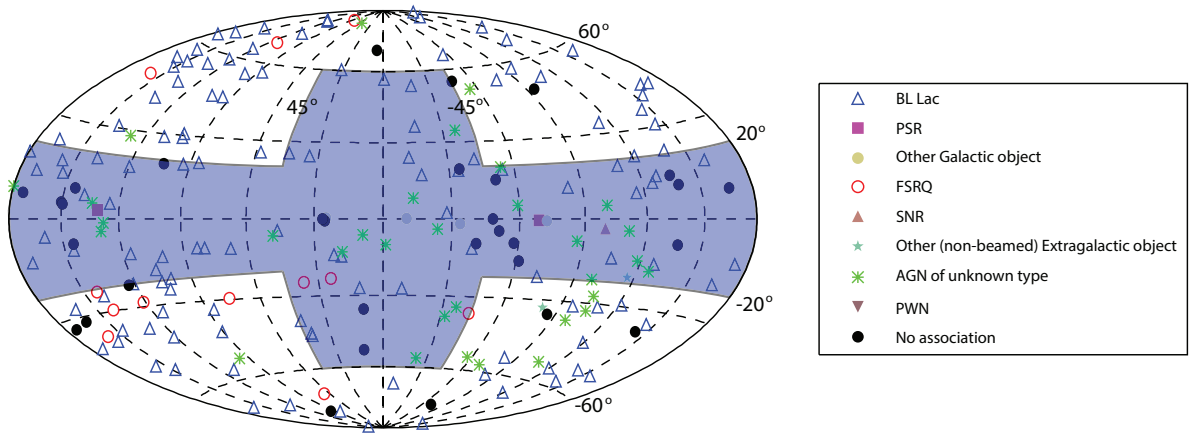}
	\caption{\small Sky map of 1FHL VHE sources \cite{SuppAckermann2013a} with the excluded region (shadowed area) for this study.}
	\label{Fig:S2}
\end{figure}

Assumptions 2 and 3 indicate that we cannot distinguish the extended halo emission from the background emission in the background bin, even if the real flux of the halo component in the background region was large. Assumptions 2 and 3 are valid if we select the data outside the region of the Galactic plane and Fermi bubbles, and the sources are isolated so that the angular distribution of photon flux in the source region is not influenced by other nearby sources. The region of the Galactic disk and Fermi bubbles is excluded to avoid anisotropic background emission, as shown in Figure \ref{Fig:S2}. Moreover, [referring to Figure \ref{Fig:S1}(b)], we exclude sources for which the distance to their nearest source, $\theta_d$, falls below some cut value, $\theta_{cut}$, which is the minimum allowed angular distance from the candidate source to its nearby sources. If assumptions 2 and 3 are satisfied for both the candidate source and its nearby source with the same values of $\theta_{in}$ and $\theta_{out}$, the ideal uncontaminated angular distribution in the source bins of a candidate source requires $\theta_d>\theta_{in}+\theta_{cut}$ and $\theta_{cut}\ge\theta_{in}$. However, this proved to be an overly restrictive criterion, resulting in very few viable sources and limited statistics in the stacked photon counts. Instead, we choose $\theta_d>\theta_{cut}\ge\theta_{in}$ to ensure that the center of the source region is not contaminated, but require an additional solid angle exclusion region for cases where nearby sources overlap the background bin. From assumptions 2 and 3, we can correct for the impact of the nearby sources by defining an exclusion region of radius $\theta_{cut}$ about these sources, and accounting for these exclusion regions by assuming that the signal and background effective area is reduced in proportion to the excluded solid angle. All data selection criteria (including bin widths) were designed prior to determining the signal in an effort to maximize statistics and minimize systematics, and to avoid extra trials.

We use this simple method of cutting close sources, because more sophisticated methods of modeling nearby sources (e.g. using likelihood analysis) have built-in assumptions about the number of nearby sources, the spectrum of these sources, and the angular distribution of $\gamma$-ray emission, introducing additional trials and other potential biases that are very difficult to accurately quantify. We carefully determine the values of $\theta_{in}$, $\theta_{out}$, and $\theta_{cut}$ to define a conservative exclusion region based on the PSF for even a very soft-spectrum source. We make an \emph{a-priori} choice of $\theta_{cut}=2.3^\circ$, $\theta_{out}=5^\circ$, and different values for $\theta_{in}$ depending on energy as summarized in Table I (see main text), which are chosen to be greater than the 95\% containment angle of the PSF in the corresponding energy range \cite{SuppIRF}. For a given source candidate, we determine its nearby sources from Fermi Source Catalogues-2FGL \cite{SuppNolan2012} and 1FHL \cite{SuppAckermann2013a}, decide whether it meets the isolated-source criteria, and, if so, cut the patches around its nearby sources within $\theta_{cut}$. The detection region of each source covers $\sim 10^\circ$ diameter in which the non-uniform exposure of the Fermi-LAT cannot be neglected. We correct the counts per unit solid angle for each source by using the Fermi exposure maps, normalized to the exposure level at the center of the detection region where the source is located. The calibrated counts are stacked, as illustrated in Figure \ref{Fig:S1}(c).

In this study, we selected Fermi-LAT weekly data from early 2008 August through early 2014 Feburary (weeks 9-296). The Pass 7 Reprocessed data in the SOURCE event class were selected with zenith angle greater than $100^\circ$, and only data observed when the spacecraft’s rocking angle below $52^\circ$ were considered. Following the spatial selection criteria above, we subsequently obtained 24 HSP BL Lacs with redshift $z<0.5$, as listed in Table \ref{Tab:S1}.

\begin{table}[htbp]
	\centering
	\caption{List of the 24 HSP BL Lacs with redshift $z<0.5$ in 1FHL source catalogue\cite{SuppAckermann2013a}.}
	\begin{tabular}{ lrrcc }
		\hline
		1FHL Source Name~~ & ~~~~RA~~~~ & ~~~~DEC~~ & ~~Associated AGN Name~~ & ~~Redshift \\
		\hline
		1FHL J0122.7+3425 & 20.682 & 34.418 & 1ES 0120+340 & 0.272 \\
		1FHL J0159.4+1048 & 29.874 & 10.811 & RX J0159.5+1047 & 0.195 \\
		1FHL J0208.7+3523 & 32.176 & 35.388 & BZB J0208+3523 & 0.318 \\
		1FHL J0213.1+2246 & 33.292 & 22.780 & MG3 J021252+2246 & 0.459 \\
		1FHL J0238.6-3117 & 39.657 & -31.284 & 1RXS J023832.6-311658 & 0.232 \\
		1FHL J0303.4-2407 & 45.868 & -24.128 & PKS 0301-243 & 0.260 \\
		1FHL J0316.3-2609 & 49.077 & -26.152 & RBS 0405 & 0.443 \\
		1FHL J0325.7-1647 & 51.437 & -16.793 & RBS 0421 & 0.291 \\
		1FHL J0449.4-4350 & 72.361 & -43.840 & PKS 0447-439 & 0.205 \\
		1FHL J0550.6-3215 & 87.669 & -32.260 & PKS 0548-322 & 0.069 \\
		1FHL J0710.5+5908 & 107.629 & 59.139 & 1H 0658+595 & 0.125 \\
		1FHL J0809.8+5217 & 122.461 & 52.294 & 1ES 0806+524 & 0.137 \\
		1FHL J1015.0+4925 & 153.773 & 49.427 & 1H 1013+498 & 0.212 \\
		1FHL J1023.6+2959 & 155.909 & 29.995 & RX J1023.6+3001 & 0.433 \\
		1FHL J1053.6+4931 & 163.403 & 49.521 & GB6 J1053+4930 & 0.140 \\
		1FHL J1058.6+5627 & 164.666 & 56.459 & TXS 1055+567 & 0.143 \\
		1FHL J1103.3-2329 & 165.846 & -23.492 & 1ES 1101-232 & 0.186 \\
		1FHL J1117.2+2013 & 169.305 & 20.227 & RBS 0958 & 0.138 \\
		1FHL J1137.0+2553 & 174.267 & 25.893 & RX J1136.8+2551 & 0.156 \\
		1FHL J1154.0-0010 & 178.525 & -0.169 & 1RXS J115404.9-001008 & 0.254 \\
		1FHL J1418.6+2539 & 214.659 & 25.658 & BZB J1417+2543 & 0.237 \\
		1FHL J1439.3+3933 & 219.835 & 39.555 & PG 1437+398 & 0.349 \\
		1FHL J1501.0+2238 & 225.275 & 22.639 & MS 1458.8+2249 & 0.235 \\
		1FHL J2322.5+3436 & 350.647 & 34.602 & TXS 2320+343 & 0.098 \\
		\hline
	\end{tabular}
	\label{Tab:S1}     
\end{table}

To test the sensitivity of our assumption of a uniform background, we compared the isotropic background model (null hypothesis, $H_0$) to the Fermi diffuse background model gll\_iem\_v05\_rev1 \cite{SuppFermiBackground} ($H_1$). To determine which hypothesis is better in describing the stacked background around the 24 BL Lacs listed in Table \ref{Tab:S1}, we evaluated the Bayes factors \cite{SuppKass1995} of $H_1$ against $H_0$ for the Fermi observations in the background bins around the 24 BL Lacs, in order to see which hypothesis the statistical evidence favors. The background bins (as depicted in Fig. \ref{Fig:S1}) were subdivided into three angular bins with equal solid angle (similar to the source bins). The number of combined photon counts in the $i$-th angular bin $N_i$ ($i=1,2,3$) is estimated by the Fermi background model $H_1$ given $\mu_it_1$, and the isotropic background model $H_0$ given $\mu_0t_0$, where $\mu_i$ ($i=1,2,3$) are the expected numbers of counts per unit exposure time calculated by the Fermi background model (with the anisotropic Fermi exposure taken into account for reproducing the observation of the background counts) in the $i$-th angular bin around the 24 BL Lacs, and $\mu_0$ is the averaged background count value per unit exposure time in each bin given by the null hypothesis. $t_1$ and $t_0$ are two free factors that play the role of the effective exposure time so that $\lambda_i=\mu_it_1$ and $\lambda_i=\mu_0t_0$ are the estimators of $N_i$ given by the hypotheses $H_1$ and $H_0$, respectively. 

The Bayes factor $B_{10}$ of $H_1$ against $H_0$ is given by
\begin{equation}
	B_{10}=\frac{\mathcal{L}_B(H_1|\{N_i\})}{\mathcal{L}_B(H_0|\{N_i\})},
	\label{Eq:S1p}
\end{equation}
where $\{N_i\}$ ($i=1,2,3$) is a set of stacked counts in the angular background bins around the source. The Bayesian likelihood function $\mathcal{L}_B$ is obtained by integrating over the model parameters. Hence, for either the null hypothesis ($k=0$) or the Fermi background model hypothesis ($k=1$), $H_k$ is given by
\begin{equation}
	\mathcal{L}_B(H_k|\{N_i\})=\int\mathrm{d}t_kP(\{N_i\}|\{\lambda_i\},H_k)\pi(t_k|H_k),
	\label{Eq:S2p}
\end{equation}
where $\{\lambda_i\}$ is a set of estimators of $\{N_i\}$ and $\pi(t_k|H_k)$ is the prior probability of $t_k$ given hypothesis $H_k$. In Eq. \ref{Eq:S2p}, the posterior density is given by the joint probability for a set of Poisson processes in the three angular bins: 
\begin{equation}
	P(\{N_i\}|\{\lambda_i\},H_j)=\prod_{i=1}^3\mathcal{P}(N_i|\lambda_i),
	\label{Eq:S3p}
\end{equation}
where $\mathcal{P}(N|\lambda)$ denotes the Poisson distribution of $\lambda$ at $N$. 

For these single parameter models, the prior density $\pi(t_j|H_j)$ can be given by the Jeffreys prior $\pi_J(t_j|H_j)$ \cite{SuppJeffreys1961}, which is proportional to the square root of the Fisher information $I(t_j)$. For the joint Poisson posterior (Eq. \ref{Eq:S3p}), the Fisher information is given by
\begin{equation}
	I(t_j)=\sum_{i=1}^3I_i(t_j),
	\label{Eq:S4p}
\end{equation}
and
\begin{equation}
	I_i(t_0)=E\left[\left(\frac{\mathrm{d}}{\mathrm{d}t_0}\mathrm{ln}\mathcal{P}(N_i|\lambda_i(t_0))\right)^2\right]=E\left[\left(\frac{N_i-\mu_0t_0}{t_0}\right)^2\right]=\frac{\mu_0}{t_0},
	\label{Eq:S5p}
\end{equation}
\begin{equation}
	I_i(t_1)=E\left[\left(\frac{\mathrm{d}}{\mathrm{d}t_1}\mathrm{ln}\mathcal{P}(N_i|\lambda_i(t_1))\right)^2\right]=E\left[\left(\frac{N_i-\mu_it_1}{t_1}\right)^2\right]=\frac{\mu_i}{t_1},
	\label{Eq:S6p}
\end{equation}
where $E[f(N|t)]$ denotes the expectation over values for $N$ with respect to the probability distribution function $f(N|t)$ for a given $t$. Hence, we choose
\begin{equation}
	\pi(t_0|H_0)=\sqrt{\frac{3\mu_0}{t_0}},
	\label{Eq:S7p}
\end{equation}
\begin{equation}
	\pi(t_1|H_1)=\sqrt{\frac{\sum_{i=1}^3{\mu_i}}{t_1}}.
	\label{Eq:S8p}
\end{equation}

The resulting Bayes factors in the four energy bins are listed in Table \ref{Tab:S2}. All the values of $\mathrm{log}_{10}B_{10}$ are less than $0.5$, showing no significant evidence in favor of the Fermi background model \cite{SuppKass1995}. This result also indicates that for the detection regions in our study, the two hypotheses are very close in describing the distributions of GeV photon events. This is understandable since anisotropies in the GeV background are mainly expected in the region of the Galaxy disk and the Fermi bubbles, which are excluded in this study, as shown in Fig. \ref{Fig:S2}. We chose not to use the Fermi diffuse background model in this study, since such models could contain contributions from the unknown extended sources that we are interested in, and would then lower the possibility of getting a detection of such extended emission.

\begin{table}[htbp]
	\centering
	\caption{Bayes factors ($B_{10}$) of the background model gll\_iem\_v05\_rev1 ($H_1$) against the uniform background hypothesis ($H_0$) for the observation in the background bins around the 24 BL Lacs.}
	\begin{tabular}{ ccccc }
		\hline
		Energy (GeV)~~ & ~~1-1.58~~ & ~~1.58-3.16~~ & ~~3.16-10~~ & ~~10-100 \\
		$\rm{log}_{10}B_{10}$ & 0.18 & -0.14 & -0.08 & 0.09 \\
		\hline
	\end{tabular}
	\label{Tab:S2}     
\end{table}

\section{\sc \large Angular Size of Pair Halos}

Following the discussion in A. Neronov and D. V. Semikoz \cite{SuppNeronov2009} (see also \cite{SuppTashiro2013}), we derive the typical angular size of a pair halo as a function of the observed energy of cascade photons $E_\gamma$, the typical redshift of the source $z_s$, and the IGMF strength $B_0$ at the present epoch, given by
\begin{equation}
	\Theta(E_\gamma,z_s,B_0)\approx 9.2\times 10^{-4}\left[1+z_{\gamma\gamma}(E_\gamma,z_s)\right]^{-2}\left(\frac{E_\gamma}{100\mathrm{GeV}}\right)^{-1}\left(\frac{B_0}{10^{-16}\mathrm{G}}\right)\left[\frac{d_\gamma(E_\gamma,z_s)}{d_s(z_s)}\right],
	\label{Eq:S1}
\end{equation}
The basic geometry of propagation of the direct and cascade $\gamma$-rays from the source to the observer (see Fig. 3 in \cite{SuppNeronov2009}) gives the typical opening angle of the cascade emission, with the small angle approximation, as
\begin{equation}
	\Theta=\frac{d_\gamma}{d_s}\delta,
	\label{Eq:S2}
\end{equation}
where $d_\gamma$ and $d_s$ are the commoving mean free path for pair production and the commoving distance to the source, respectively, and $\delta$ is the deflection angle of electron-positron pairs by the IGMF. Assuming that the correlation length of the IGMF $\lambda_B$ is much greater than the mean free path for IC scattering $D_e$, $\delta$ can be estimated as the ratio of $D_e$ and the Larmor radius of electron $R_L$ in the magnetic field, which are given by
\begin{equation}
	D_e\approx 10^{21}(1+z_{\gamma\gamma})^{-4}\left(\frac{E_e}{10\mathrm{TeV}}\right)^{-1}\mathrm{m}\approx 2\times 10^{21}(1+z_{\gamma\gamma})^{-4}\left(\frac{E_{\gamma_0}}{10\mathrm{TeV}}\right)^{-1}\mathrm{m},
	\label{Eq:S3}
\end{equation}
\begin{equation}
	R_L\approx \frac{E_e}{ceB}\approx \frac{E_{\gamma_0}}{2ceB_0(1+z_{\gamma\gamma})^2}\approx 1.67\times 10^{24}(1+z_{\gamma\gamma})^{-2}\left(\frac{E_{\gamma_0}}{10\mathrm{TeV}}\right)\left(\frac{B_0}{10^{-16}\mathrm{G}}\right)^{-1}\mathrm{m}.
	\label{Eq:S4}
\end{equation}
where we assumed that the pair produced electron/positron has half the energy of the initial photon, $E_e\approx E_{\gamma_0}/2$. We related the magnetic field at the time of the pair production and IC scattering, $B$, to the magnetic field today, $B_0$, assuming that it is only affected by the $(1+z_{\gamma\gamma})^{-2}$ redshift dilution. For $E_{\gamma_0}\approx10$ TeV, $E_e\epsilon_{\mathrm{CMB}}\approx3\times10^9\mathrm{eV}\ll(m_ec^2)^2\approx2.5\times10^{11}\mathrm{eV}$, where $\epsilon_{\mathrm{CMB}}$ is the typical energy of a CMB photon. Hence, we can use the Thomson approximation where the energy of cascade $\gamma$-rays produced by IC scattering, as observed on Earth, is given by
\begin{equation}
	E_\gamma=\frac{4}{3}(1+z_{\gamma\gamma})^{-1}\epsilon_{\mathrm{CMB}}'\left(\frac{E_e}{m_ec^2}\right)^2.
	\label{Eq:S5}
\end{equation}
Inserting the typical energy of a CMB photon at redshift $z_{\gamma\gamma}$, when the IC occurs, $\epsilon_{\mathrm{CMB}}'=6\times 10^{-4}(1+z_{\gamma\gamma})\mathrm{eV}$, we obtain
\begin{equation}
	E_\gamma=77\mathrm{GeV}\left(\frac{E_{\gamma_0}}{10\mathrm{TeV}}\right)^2.
	\label{Eq:S6}
\end{equation}
Hence, the deflection angle is given by
\begin{equation}
	\delta=\frac{D_e}{R_L}\approx 9.2\times 10^{-4}(1+z_{\gamma\gamma})^{-2}\left(\frac{E_\gamma}{100\mathrm{GeV}}\right)^{-1}\left(\frac{B_0}{10^{-16}\mathrm{G}}\right).
	\label{Eq:S7}
\end{equation}
Eq. \ref{Eq:S1} can be obtained by substituting Eq. \ref{Eq:S7} into Eq. \ref{Eq:S2}.

Assuming a spatially flat Friedmann-Lema\^{\i}tre-Robertson-Walker (FLRW) universe, if a TeV photon is emitted at time $t_s$ from the AGN and pair produces on an EBL photon at time $t_{\gamma\gamma}$, the commoving distances $d_\gamma$ and $d_s$ are given by
\begin{equation}
	d_\gamma=\int_{t_s}^{t_{\gamma\gamma}}c\frac{\mathrm{d}t}{a(t)},
	\label{Eq:S8}
\end{equation}
\begin{equation}
	d_s=\int_{t_s}^{t_{\mathrm{now}}}c\frac{\mathrm{d}t}{a(t)},
	\label{Eq:S9}
\end{equation}
where $a(t)$ is the scale factor at time $t$. Using the change of variables
\begin{equation}
	\frac{\mathrm{d}t}{\mathrm{d}z}=-\frac{a_0}{H(1+z)},
	\label{Eq:S10}
\end{equation}
and $a(z)=a_0/(1+z)$, we can express the comoving distance (Eq. \ref{Eq:S8}, \ref{Eq:S9}) in terms of the redshift of emission and pair production, $z_s$ and $z_{\gamma\gamma}$. Hence,
\begin{equation}
	d_\gamma=\frac{c}{H_0}\int_{z_{\gamma\gamma}}^{z_s}\frac{\mathrm{d}z}{\sqrt{\Omega_M(1+z)^3+\Omega_\Lambda}},
	\label{Eq:S11}
\end{equation}
\begin{equation}
	d_s=\frac{c}{H_0}\int_{0}^{z_s}\frac{\mathrm{d}z}{\sqrt{\Omega_M(1+z)^3+\Omega_\Lambda}},
	\label{Eq:S12}
\end{equation}
For the redshifts of interest, we can take the universe as made of matter and cosmological constant only, and the Hubble parameter is given by $H=H_0\sqrt{\Omega_M(1+z)^3+\Omega_\Lambda}$ for a flat matter-$\Lambda$ FLRW universe.

Note that $z_{\gamma\gamma}$ in Eq. \ref{Eq:S1} and Eq. \ref{Eq:S11} cannot be measured directly. Taking the expression of the mean free path (which assumes a redshift dependence of the EBL number density $\propto(1+z)^{-2}$) from \cite{SuppNeronov2009}:
\begin{equation}
	D_\gamma\approx 80\frac{\kappa}{(1+z)^2}\left(\frac{E_{\gamma_0}}{10\mathrm{TeV}}\right)^{-1}\mathrm{Mpc}\approx 80\frac{\kappa}{(1+z)^2}\left(\frac{E_\gamma}{77\mathrm{GeV}}\right)^{-1/2}\mathrm{Mpc},
	\label{Eq:S13}
\end{equation}
where $\kappa\sim 1$ accounts for the EBL model uncertainties, and Eq. \ref{Eq:S6} is used to express the mean free path $D_\gamma$ in terms of $E_\gamma$. The optical depth of the $\gamma$-ray propagating from the source grows as
\begin{equation}
	\frac{\mathrm{d}\tau}{\mathrm{d}t}=\frac{c}{D_\gamma(E_\gamma,z)}.
	\label{Eq:S14}
\end{equation}
The time of pair production corresponds to when $\tau$ reaches 1, and can be found implicitly from
\begin{equation}
	\int_{t_s}^{t_{\gamma\gamma}}\frac{c\mathrm{d}t}{D_\gamma(E_\gamma,z)}=\int_{z_s}^{z_{\gamma\gamma}}\frac{c\mathrm{d}z}{D_\gamma(E_\gamma,z)}\frac{\mathrm{d}t}{\mathrm{d}z}=1.
	\label{Eq:S15}
\end{equation}
Eq. \ref{Eq:S15} allows us to solve for $z_{\gamma\gamma}(E_\gamma,z_s)$ implicitly in terms of $E_\gamma$ and $z_s$. We can then obtain $d_\gamma(z_{\gamma\gamma},z_s)$ and $d_s(z_s)$ from Eq. \ref{Eq:S11} and Eq. \ref{Eq:S12}, and find the angular size of pair halos $\Theta$ as a function of $E_\gamma$, $z_s$, and $B_0$ from Eq. \ref{Eq:S1}.

For large $z_s$ and $E_\gamma$, one can assume that $z_{\gamma\gamma}\approx z_s$ \cite{SuppNeronov2009,SuppTashiro2013}, leading to
\begin{equation}
	\Theta\approx 1.5\times 10^{-5}(1+z_s)^{-3}\left(\frac{E_\gamma}{100\mathrm{GeV}}\right)^{-3/2}\left(\frac{B_0}{10^{-16}\mathrm{G}}\right)\left(\int_{0}^{z_s}\frac{\mathrm{d}z}{\sqrt{\Omega_M(1+z)^3+\Omega_\Lambda}}\right)^{-1}.
	\label{Eq:S16}
\end{equation}
However, this assumption is not true in general. In particular, it overestimates the angular extent of the pair halos around low redshift sources. Hence, we do not make this assumption when using Eq. \ref{Eq:S1} to estimate $B_0$ from the most likely values of $\Theta$.

\begin{figure}[htbp!]
	\centering
	\includegraphics[width=13cm, angle=0]{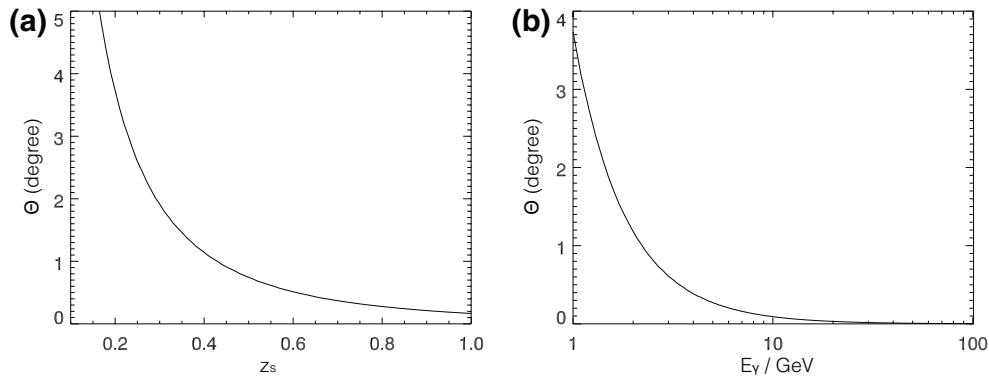}
	\caption{\small Redshift and energy dependence of the angular size of pair halos produced by IGMF of strength $B_0=10^{-16}\mathrm{G}$, given by Eq. \ref{Eq:S1}. (a) The $z_s$ dependence of the typical angular size $\Theta$ of pair halos at $E_\gamma=1\mathrm{GeV}$. (b) The $E_\gamma$ dependence of the typical angular size $\Theta$ of pair halos from AGNs with redshift $z_s=0.2$.}
	\label{Fig:S3}
\end{figure}

From Eq. \ref{Eq:S1}, it is obvious that $\Theta\propto B_0$. The $z_s$ and $E_\gamma$ dependence of $\Theta$, however, is not explicit. Figure \ref{Fig:S3} shows the sensitivity of $\Theta$ to various model parameters assuming an IGMF of $B_0=10^{-16}\mathrm{G}$ and using Eq. \ref{Eq:S1}: The $z_s$ dependence of $\Theta$ for $E_\gamma=1\mathrm{GeV}$ is shown in Figure \ref{Fig:S3}(a), and the $E_\gamma$ dependence of $\Theta$ for $z_s=0.2$ is shown in Figure \ref{Fig:S3}(b). From Figure \ref{Fig:S3}(a), we find it is quite unlikely to detect the extended emission from high-redshift sources, supporting our selection criteria for stacking sources based on redshift. We can also find, from Figure \ref{Fig:S3}(b), that lower energy electrons are deflected by larger angles, consistent with the results we have obtained in this study (as discussed in the main text). 

\section{\sc \large Details of the stacking source analysis}

\begin{figure}[htbp!]
	\centering
	\includegraphics[width=14cm, angle=0]{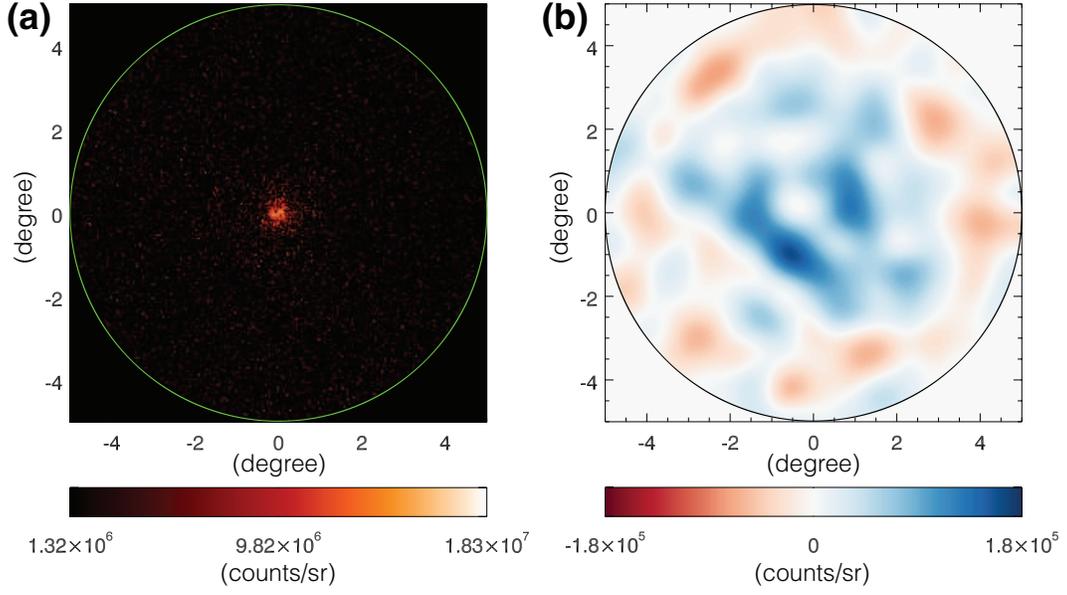}
	\caption{\label{Fig:1s}$\gamma$-ray counts maps of the stacked sources in the 1GeV-1.58GeV energy bin. The large circles show the outer edge of the detection region. (a) Counts map of the stacked BL Lacs. (b) Smoothed counts difference between the stacked BL Lacs and the center-normalized stacked FSRQs. Positive values indicate the BL Lacs' counts are greater than the normalized counts of the FSRQs in that angular region}.
\end{figure}

We identify 24 HSP BL Lacs with redshift $z<0.5$ that satisfy our selection criteria and we stack their photon events, as shown in Fig. \ref{Fig:1s}(a) ($\gamma$-ray counts map in 1 GeV-1.58 GeV). As a control population, 26 FSRQs (with any redshift) are also selected by the same criteria. Fig. \ref{Fig:1s}(b) shows the difference of the $\gamma$-ray counts in 1 GeV-1.58 GeV between the two source populations. The background counts of these two stacked sources calculated by averaging the counts in the background bin are then subtracted from their total counts. To make the two populations comparable, the background-subtracted counts of the stacked FSRQs are normalized to the same level as that of the stacked BL Lacs at the center. We smooth the counts maps by using a Gaussian kernel with full width at half maximum of $1^\circ$, and subtract the normalized FSRQs' counts from the BL Lacs'. In Fig. \ref{Fig:1s}(b), the difference map shows an excess of the $\gamma$-ray emission around the stacked BL Lacs over the stacked FSRQs.

\begin{figure}[htbp!]
	\centering
	\includegraphics[width=16cm, angle=0]{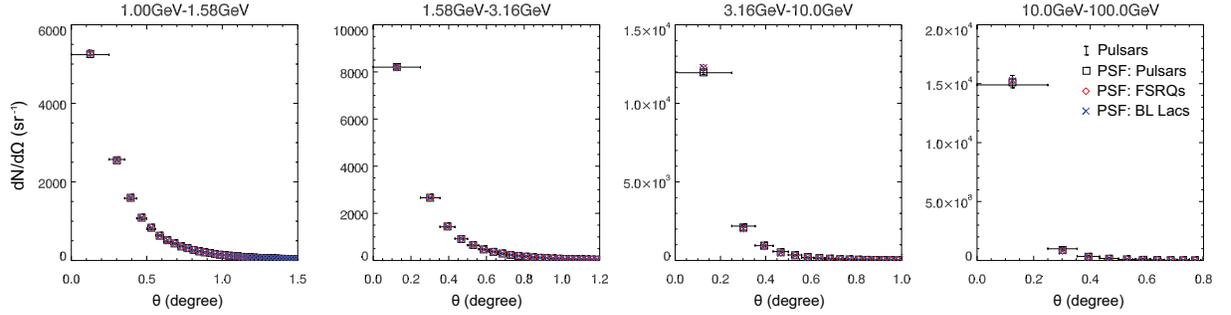}
	\caption{\small Angular distribution of photon events around the stacked pulsars (error bars): vertical errors are the 68\% confidence intervals, horizontal errors show the size of angular bins; Angular distribution of effective PSFs calculated for the stacked pulsars (squares), BL Lacs (crosses), and FSRQs (diamonds) for the same observation times and the observed spectrum.}
	\label{Fig:S4}
\end{figure}

\begin{figure}[htbp!]
	\centering
	\includegraphics[width=16cm, angle=0]{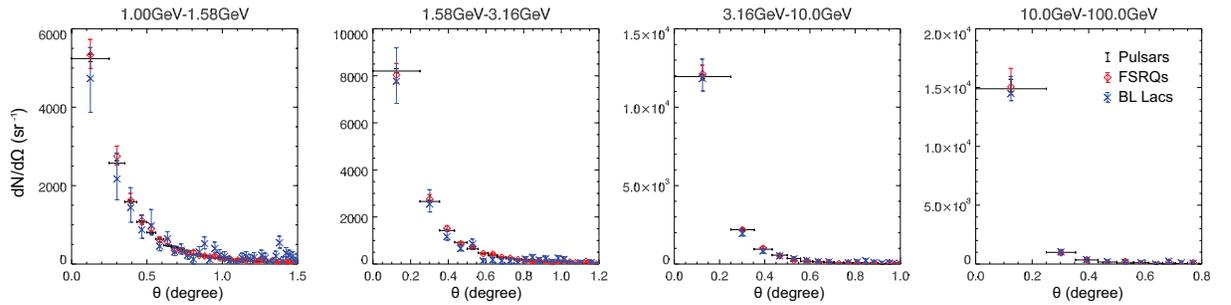}
	\caption{\small Angular distribution of photon events around the stacked pulsars (black), the stacked FSRQs (red), and the stacked BL lacs (blue): vertical errors are the 68\% confidence intervals; horizontal errors show the size of angular bins.}
	\label{Fig:S4l}
\end{figure}

We choose the Crab and Geminga pulsars as our calibration sources since they are effective point sources for Fermi-LAT \cite{SuppNeronov2011,SuppAckermann2013}. Figure \ref{Fig:S4} shows the angular distribution of photon events around the stacked pulsars and the angular distribution of the effective PSFs calculated for the same observation times and the observed spectrum. We use the Fermi Science Tools to calculate the PSF for the same observational parameters as our different data sets. We also plot these calculated PSF profiles for the 24 stacked BL Lacs and the 26 stacked FSRQs in the same figure. There is only a very slight difference among their PSFs, hence the normalized stacked profiles of the three set of sources are roughly comparable. The good consistency in the angular distributions determined with both the pulsar and PSF data sets leave the appearance of extended emission about the BL Lac data set (as shown in Fig. \ref{Fig:1s}) the notable exception. We calculate the normalized angular profiles for the stacked pulsars, the 24 BL Lacs, and the 26 FSRQs, as shown in Fig. \ref{Fig:S4l} (the same profiles as shown in Fig. 1 in the main text, including the other two higher energy bins). Again, in the lower energy bins, the normalized angular profiles of stacked BL Lacs have lower scaled counts per unit solid angle at small $\theta$, providing evidence for extended emission since the additional counts in the extended halo reduce the scaled counts at small angles after normalization.

\section{\sc \large Details of the frequentist analysis}

In our study, we evaluate the likelihood in the two-dimensional space of $f_{\mathrm{halo}}$ and $\Theta$. As described in the main text, the model hypothesis $\mathcal{H}_1$ is defined on a multidimensional space of model parameters $\bm{x}\equiv\{f_{\mathrm{halo}},\Theta,\bm{\mu},\bm{A}\}$, where $\bm{\mu}$ and $\bm{A}$ are a set of background values (in counts per unit solid angle) and a set of normalization factors, respectively, for a group of stacked sources. We define $h_1(f_{\mathrm{halo}},\Theta)$ as a subset of $\mathcal{H}_1$  for a given pair of $f_{\mathrm{halo}}$ and $\Theta$. $\mathcal{H}_0$ corresponds to the null hypothesis with $\bm{x}$ constrained by $f_{\mathrm{halo}}=0$ or $\Theta=0$. The values of the likelihood ratio $\Lambda(f_{\mathrm{halo}},\Theta|\bm{N})=\mathrm{sup}\{\mathcal{L}(h_1|\bm{N}):\bm{x}\in h_1\}/\mathrm{sup}\{\mathcal{L}(\mathcal{H}_0|\bm{N}):\bm{x}\in \mathcal{H}_0\}$ are evaluated and displayed in two-dimensional ($f_{\mathrm{halo}},\Theta$)-space, where $\bm{N}$ denotes the set of observations $\{N_{i,j}\}$ (see main text), and $sup$ is the supremum function. As discussed in the main text, since $N_{i,j}$ is a very small number, a direct frequentist approach where one calculates the joint likelihood
\begin{equation}
	L\equiv\prod_{i,j}\mathcal{P}(N_{i,j}|\lambda_{i,j})
	\label{Eq:S16-0}
\end{equation}
would lead to a non-converging test statistic (TS) distribution and potentially result in a Type II error \cite{SuppWilks1938,SuppBaldwin2013}. To overcome the problem of small sample size, we repartition the data into two sets: the stacked angular distribution $\{\sum_{i=1}^{n}N_{i,j}\}\equiv\{\eta_j\}$ obtained by summing over sources $i$, and the stacked source distribution $\{\sum_{j=1}^{m}N_{i,j}\}\equiv\{\zeta_i\}$ obtained by summing over angular bins $j$, where $m$ and $n$ are the total number of angular bins and stacked sources, respectively. 

Given $\mathcal{N}$ number of samples, the probability of partitioning the samples into $k$ parts with $\{n_1,n_2,...,n_k\}$ samples in each part follows a multinomial distribution
\begin{equation}
	\mathcal{M}_{p_1,p_2,...,p_k}^\mathcal{N}(n_1,n_2,...,n_k)=\frac{\mathcal{N}!}{n_1!n_2!\cdots n_k!}p_1^{n_1}p_2^{n_2}\cdots p_k^{n_k},
	\label{Eq:S16-1}
\end{equation}
where $p_1,p_2,...,p_k$ are the probabilities giving $n_1,n_2,...,n_k$ in each part, respectively. Hence, the likelihood of obtaining $\{\zeta_i\}$ and $\{\eta_j\}$, $\mathcal{L}_{\rm{on}}$, is the probability of having a $N_{\mathrm{tot}}\equiv\sum_{i,j}N_{i,j}$ total counts with the two independent ways of repartitioning the data given by $\{\zeta_i\}$ and $\{\eta_j\}$, respectively. Thus we have
\begin{equation}
	\mathcal{L}_{\mathrm{on}}(\bm{x}|\{\zeta_i\},\{\eta_j\})=\mathcal{P}(N_{\mathrm{tot}}|\lambda_{\mathrm{tot}})\mathcal{M}_{\{p_{\zeta,i}\}}^{N_{\mathrm{tot}}}{\{\zeta_i\}}\mathcal{M}_{\{p_{\eta,j}\}}^{N_{\mathrm{tot}}}{\{\eta_j\}}.
	\label{Eq:S16-2}
\end{equation}
For a given set of model parameters $\bm{x}=(f_{\mathrm{halo}},\Theta,\bm{\mu},\bm{A})$, the estimators $\lambda_{\mathrm{tot}}\equiv\sum_{i,j}\lambda_{i,j}$, $p_{\zeta,i}\equiv\sum_j\lambda_{i,j}/\lambda_{\mathrm{tot}}$ and $p_{\eta,j}\equiv\sum_i\lambda_{i,j}/\lambda_{\mathrm{tot}}$ can be calculated using Eq. 3 in the main text. Note that we can always rewrite a joint Poisson distribution (e.g. Eq. \ref{Eq:S16-0}) as the product of a Poisson distribution and a multinomial distribution
\begin{equation}
	\prod_{i=1}^k\mathcal{P}(n_i|\lambda_i)=\mathcal{P}(\mathcal{N}|\lambda)\mathcal{M}_{p_1,p_2,...,p_k}^\mathcal{N}(n_1,n_2,...,n_k),
	\label{Eq:S16-3}
\end{equation}
where $\mathcal{N}=\sum_{i=1}^kn_i$, $\lambda=\sum_{i=1}^k\lambda_i$, and $p_i=\lambda_i/\lambda$.

This likelihood $\mathcal{L}_{\mathrm{on}}$ is combined with the likelihood of getting a set of $\{N_{i,m}\}$ counts detected in each background bin around each source $\mathcal{L}_{\rm{off}}$:
\begin{equation}
	\mathcal{L}_{\mathrm{off}}(\bm{x}|\{N_{i,m}\})=\prod_{i=1}^n\mathcal{P}(N_{i,m}|\lambda_{i,m}).
	\label{Eq:S16-4}
\end{equation}

\begin{figure}[htbp]
	\centering
	\includegraphics[width=16cm, angle=0]{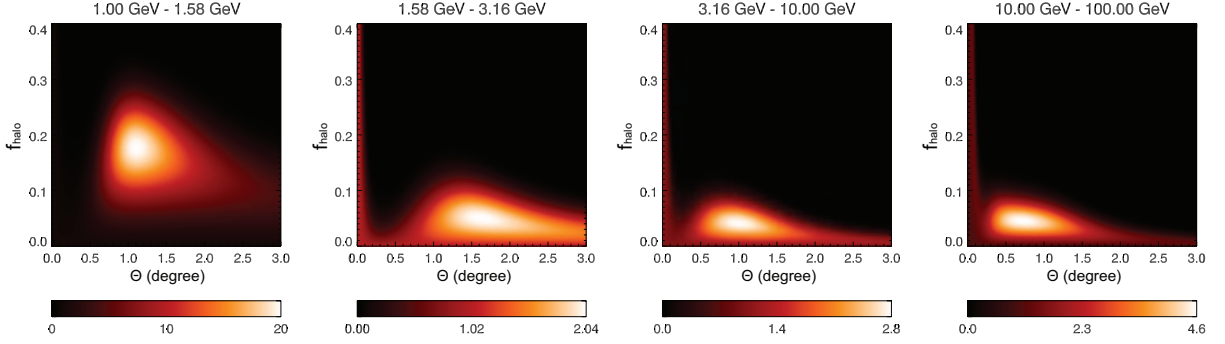}
	\caption{\small Likelihood ratio maps for stacked BL Lacs. Colors show the ratio of the likelihood of extended-emission hypothesis to that of the null hypothesis (the PSF).}
	\label{Fig:S5}
\end{figure}

\begin{figure}[htbp]
	\centering
	\includegraphics[width=16cm, angle=0]{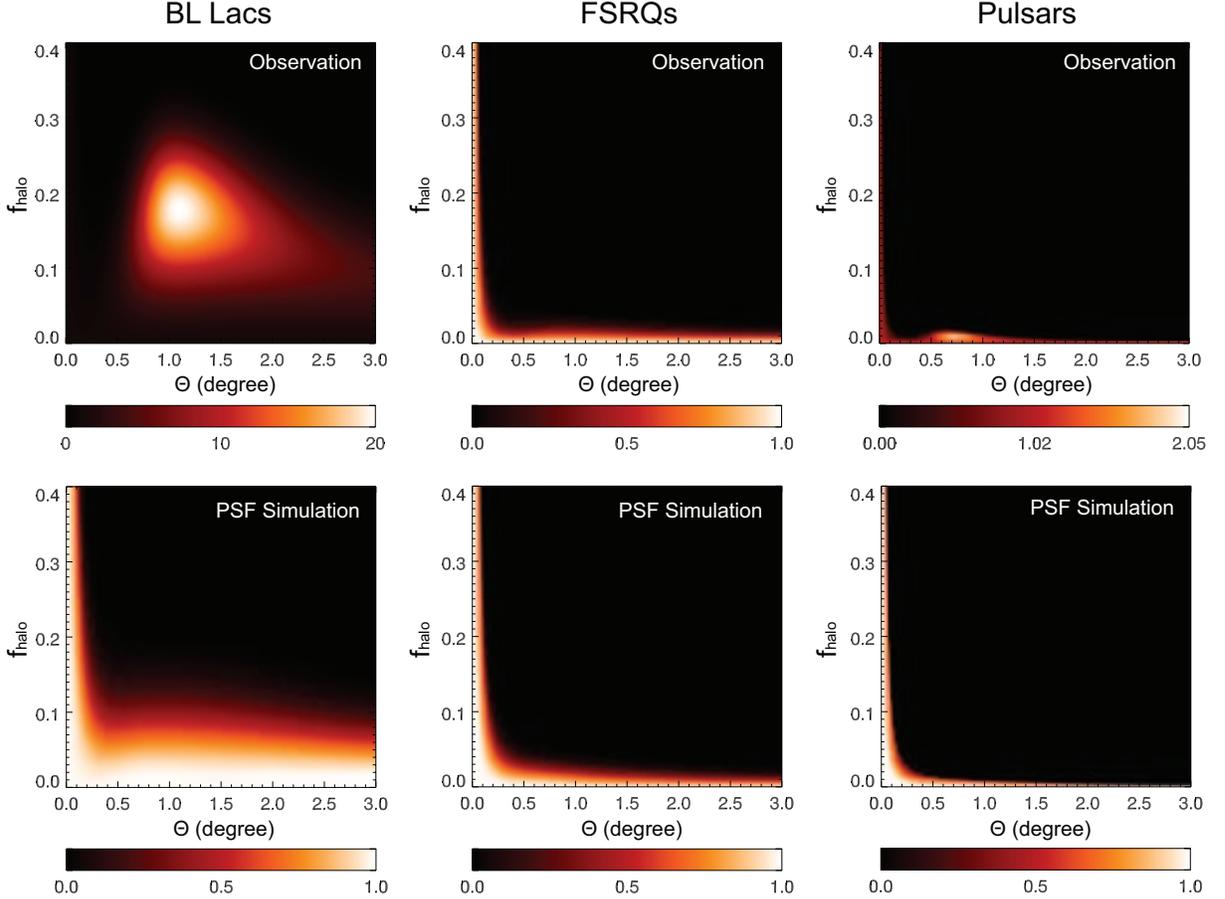}
	\caption{\small Likelihood ratio maps in the 1-1.58 GeV energy bin. Colors show the ratio of the likelihood of extended-emission hypothesis to that of the null hypothesis (the PSF). Top three panels show likelihood ratio maps for stacked BL Lacs, FSRQs, and pulsars; Bottom three panels show the likelihood ratio maps for point sources with angular distribution given by the PSF with total number of events set equal to that of the corresponding stacked BL Lacs, FSRQs, and pulsars.}
	\label{Fig:S6}
\end{figure}

We subsequently evaluate the joint likelihood $\mathcal{L}=\mathcal{L}_{\rm{on}}\times\mathcal{L}_{\rm{off}}$ where $\zeta_i$, $\eta_j$, and $N_{i,m}$ are all relatively large numbers of counts, as discussed in the main text. The supremum likelihood value for a given $f_{\mathrm{halo}}$ and $\Theta$ is found in the $2n$-dimensional space of model parameters $\bm{\mu}$ and $\bm{A}$ (where $n=2$ for the stacked pulsars, $n=24$ for the BL Lacs, and $n=26$ for the FSRQs) using Powell's method \cite{SuppPowell1964}. The resulting likelihood ratio maps (Fig. \ref{Fig:S5}) show peaks at non-zero $f_{\mathrm{halo}}$ and $\Theta$ in the first energy bins, while the likelihood maps for the higher energy bins peak close to the axes, consistent with the null hypothesis. This matches our expectation based on the decreasing angular scale of the halo for increasing energy as seen in Eq. \ref{Eq:S1}. So we do not include an additional trials factor for looking in these different energy bins. Furthermore, we calculate the likelihood maps for the simulated point source (labelled PSF) with a total number of events in each energy bin set to that of the stacked BL Lacs in the first energy bin. Whenever calculating the simulated point-source maps we use the instrument response parameters for the corresponding observing time and source position obtained by using the Fermi Science Tools. The likelihood maps for the stacked FSRQs, pulsars, together with their corresponding maps for simulated point sources with the same number of counts, as shown in Figure \ref{Fig:S6}. From the distributions of the maximum likelihood and the values of the likelihood ratio, we can find significant difference between the observed BL Lacs and the simulated point source, while the FSRQs and pulsars are shown to appear as point sources for Fermi-LAT.

\begin{figure}[htbp!]
	\centering
	\includegraphics[width=7cm, angle=0]{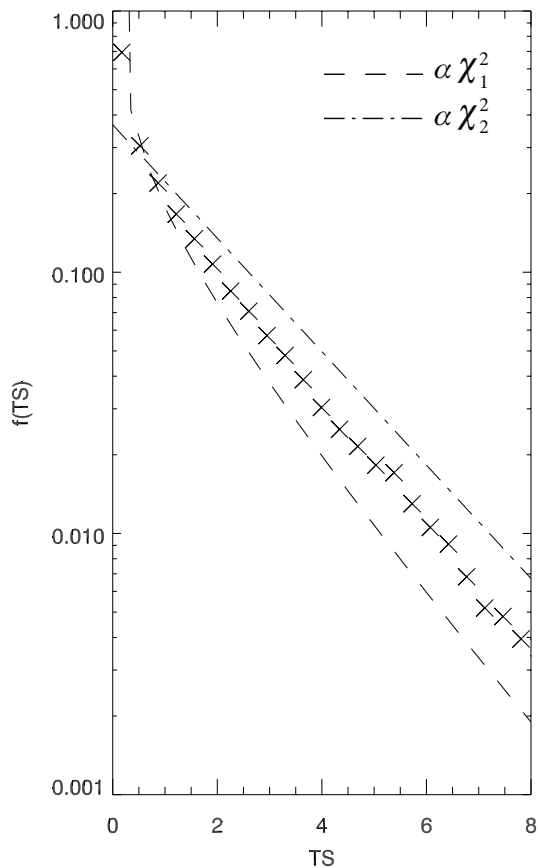}
	\caption{\small Probability distribution of the test statistic (TS), as shown by the crosses. Dashed line and the dot-dashed line are the chi-squared distributions, bounding the points (crosses) determined by Monte Carlo calculation.}
	\label{Fig:S7}
\end{figure}

A classical likelihood ratio test (LRT) applied to this problem is potentially inaccurate since the probability distribution of the test statistic (TS) is non-trivial. Wilks' theorem gives a useful approximation: the distribution of a $\mathrm{TS}=2\mathrm{ln}\Lambda$ (the likelihood ratio $\Lambda$ as defined in the main text) for nested hypotheses will be asymptotically $\chi^2$-distributed as the sample size goes to infinity. However, the theorem is only valid under certain conditions including restrictions on the sample population and the formulation of the hypotheses to be tested \cite{SuppProtassov2002,SuppMattox1996}. In this study, the set of the model parameters $\boldsymbol{x}=(f_{\mathrm{halo}},\Theta,\bm{\mu},\bm{A})$ is not open and the null hypothesis is defined on the boundaries of the domain. In such a case, we cannot directly apply Wilks' theorem to determine the distribution of the TS \cite{SuppProtassov2002}. In the study of \cite{SuppMattox1996}, a Monte Carlo (MC) method is used to check the distribution of TS, and a multiplicative factor of $\alpha=0.5$ is found in the resulting $\chi^2$-distribution because the null hypothesis stands on the symmetric boundaries of the parameter space, which indicates that half of the MC samples give positive TS values following a $\chi^2$-distribution, while the rest of the MC samples maximize the likelihood under the null hypothesis, giving $\mathrm{TS}=0$. We also apply the MC method to determine the distribution of the TS. We find that the probability distribution of the non-zero TS values lays between the distributions of $\chi^2_1$ and $\chi^2_2$, and the ratio of non-zero TS values to the total MC samples $\alpha\approx0.73$ given by the MC simulation, as shown in Figure \ref{Fig:S7}. In our study, the maximum likelihood ratio gives a TS of $\sim 6$ (as shown in Fig. 2 in the main text), corresponding to a p-value of $\sim 0.01$, as shown in Fig. \ref{Fig:S7}. This p-value indicates a significance which is equivalent to the probability of getting a normal-distributed sample $x\sim \mathcal{N}(\mu,\sigma^2)$ for $|x-\mu|>n\sigma$ with $n \sim 2.3$ (i.e. $\sim2.3\sigma$ significance). We emphasize that this method, based on repartitioned data, uses only measured (not weighted) counts for which the probability distribution function is known exactly. While this method gives good convergence (in finding the maximum likelihood value), it does so with the loss of some information. The following Bayesian analysis keeps all information and results in the conclusion that the data reveal somewhat stronger evidence in support of the pair-halo hypothesis. 

\section{\sc \large Bayes Factors of the Pair Halo Detection}

In the above frequentist analysis, we cannot provide an analytical function for the distribution of the TS values. And we have to combine the counts into different sets to overcome the problem of small sample size. Alternatively, as suggested by \cite{SuppProtassov2002}, we can test the statistical significance of the extended emission by evaluating the Bayes factors \cite{SuppKass1995}, which can well handle the small sample size problem, but needs much more computational time \cite{SuppBaldwin2013}. 

We introduce a hypothesis of extended emission for a given $f^*_{\mathrm{halo}}$ and $\Theta^*$, $H_1(f^*_{\mathrm{halo}},\Theta^*)$, which is defined as a subset of $\mathcal{H}_1$ (see the main text) for $f_{\mathrm{halo}}=f^*_{\mathrm{halo}}$ and $\Theta=\Theta^*$. The Bayes factors of $H_1(f^*_{\mathrm{halo}}, \Theta^*)$ against the null hypothesis $\mathcal{H}_0$ are given by:
\begin{equation}
	B_{10}(f^*_{\mathrm{halo}}, \Theta^*)=\frac{\mathcal{L}_B(H_1|\boldsymbol{N})}{\mathcal{L}_B(\mathcal{H}_0|\boldsymbol{N})},
	\label{Eq:S17}
\end{equation}
where $\mathcal{L}_B$ is the Bayesian likelihood function. Applying Bayes’ theorem, the likelihood function is given by the Bayesian probability, which is obtained by integrating (not maximizing) over the parameter space \cite{SuppKass1995}. Hence, for a hypothesis $H_k(H_k=\mathcal{H}_0,H_1)$,
\begin{equation}
	\mathcal{L}_B(H_k|\boldsymbol{N})=\int\mathrm{d}\boldsymbol{x}P(\boldsymbol{N}|\boldsymbol{\lambda}(\boldsymbol{x}),H_k)\pi(\boldsymbol{x}|H_k),
	\label{Eq:S18}
\end{equation}
where $\boldsymbol{N}$ is the set of observed counts $\{N_{i,j}\}$ in the angular bins, $\boldsymbol{x}=(f_{\mathrm{halo}},\Theta,\bm{\mu},\bm{A})$ is the set of model parameters, and $\boldsymbol{\lambda}$ is the set of Poisson estimators $\{\lambda_{i,j}\}$ given by the halo model. For a Bayesian method, the problem of limited statistics in the ($i$, $j$) bins is eliminated \cite{SuppBaldwin2013}, the posterior density in Eq. \ref{Eq:S18} can be straightly given by Eq. \ref{Eq:S16-0}.

The prior density $\pi(\boldsymbol{x}|H_k)$ is the probability density for getting a set of model parameters $\boldsymbol{x}$ with a hypothesis $H_k$. It can be assigned by using the probability of getting the total number of counts in all the source bins ($N_{\mathrm{on},i}\equiv\sum_{j=1}^{m-1}N_{i,j}$) and in the background bin ($N_{\mathrm{off},i}\equiv N_{i,m}$) around each source $i$. These measurements can help us to evaluate $\pi(\boldsymbol{x}|H_k)$ if we assume that they are prior measurements which can provide us the knowledge of how the data from the stacked sources are combined, revealing the prior information of the model parameters $\bm{\mu}$ and $\bm{A}$. The measurements of $N_{\mathrm{on},i}$ and $N_{\mathrm{off},i}$ can also be treated as Poisson experiments. Hence,
\begin{equation}
	\pi(\boldsymbol{x}|H_k)=\prod_{i=1}^n\left[\mathcal{P}\left(N_{\mathrm{on},i}\bigg|\sum_{j=1}^{m-1}\lambda_{i,j}\right)\mathcal{P}(N_{\mathrm{off},i}|\lambda_{i,m})\right]\times\delta(f_{\mathrm{halo}}-f^*_{\mathrm{halo}})\delta(\Theta-\Theta^*),
	\label{Eq:S19}
\end{equation}
From Eq. \ref{Eq:S16-0}, \ref{Eq:S18}, and \ref{Eq:S19}, the Bayesian likelihood (Eq. \ref{Eq:S18}) can be rewritten as
\begin{equation}
	\begin{aligned}
		\mathcal{L}_B(H_k|\boldsymbol{N})=\int_0^1\mathrm{d}f_{\mathrm{halo}}\int_0^\pi\mathrm{d}\Theta\int_0^\infty\mathrm{d}\mu_1\cdots\int_0^\infty\mathrm{d}\mu_n\int_0^\infty\mathrm{d}A_1\cdots\int_0^\infty\mathrm{d}A_n\prod_{i=1}^n\left[\prod_{j=1}^m\mathcal{P}(N_{i,j}|\lambda_{i,j})\right]\\
		\times \prod_{i=1}^n\left[\mathcal{P}\left(N_{\mathrm{on},i}\bigg|\sum_{j=1}^{m-1}\lambda_{i,j}\right)\mathcal{P}(N_{\mathrm{off},i}|\lambda_{i,m})\right]\times\delta(f_{\mathrm{halo}}-f^*_{\mathrm{halo}})\delta(\Theta-\Theta^*).
	\end{aligned}
	\label{Eq:S20}
\end{equation}
$\mathcal{L}_B(\mathcal{H}_0|\boldsymbol{N})$ is then just a special case of Eq. \ref{Eq:S20} when $f^*_{\mathrm{halo}}=0$ or $\Theta^*=0$.

The Bayes factors can be consequently obtained via evaluating this multi-dimensional integral. In the 1 GeV-1.58 GeV energy bin, there are a number of hypotheses $H_1(f^*_{\mathrm{halo}},\Theta^*)$ that show evidence against the null hypothesis $\mathcal{H}_0$. Recalling the interpretation of $B_{10}$ in half-units on the $\mathrm{log}_{10}$ scale, the $1/2<\mathrm{log}_{10}B_{10}<1$, $1<\mathrm{log}_{10}B_{10}<2$, and $\mathrm{log}_{10}B_{10}>2$ provide substantial, strong, or decisive evidence against the null hypothesis, respectively \cite{SuppJeffreys1961,SuppKass1995}. We plot the contours of such levels of $\mathrm{log}_{10}B_{10}$ in the $f^*_{\mathrm{halo}}$-$\Theta^*$ coordinates [as shown in Fig. 3(a) in the main text]. We can see decisive evidence for non-zero $f^*_\mathrm{halo}$ and $\Theta^*$. For the higher energy bins, there is no hypothesis $H_1(f^*_{\mathrm{halo}},\Theta^*)$ giving substantial evidence against $\mathcal{H}_0$ ($\mathrm{log}_{10}B_{10}>0.5$), agreeing with the results shown in the frequentist analysis. The information about the IGMF is contained in the extended emission. Here we focus on the model factor $\Theta$, and seek to get the quantitative significant range of its values for the stacked BL Lacs. We introduce a hypothesis $\hat{H}_1$ for a given $\Theta^*$ with all possible values of $f_{\mathrm{halo}}$. Although the prior distribution of $f_{\mathrm{halo}}$ is very hard to determine, one can assume that the marginal density (the whole Bayesian integrand) is dominated by the posterior density (this will usually be the case for large samples \cite{SuppKass1995}), so that we do not pre-assume any knowledge of $f_{\mathrm{halo}}$ before doing the experiment. In that case, the Bayes factors of $\hat{H}_1$ can be evaluated by integrating the Bayesian likelihood $\mathcal{L}_B$ over all possible values of $f_{\mathrm{halo}}$. Thus, the resulting Bayes factors $\hat{B}_{10}$ of $\hat{H}_1$ against $\mathcal{H}_0$ are given as a function of $\Theta^*$ [as shown in Fig. 3(b) in the main text]. Summarizing, there is decisive evidence in the 1 GeV-1.58 GeV energy bin in favor of the pair halos with angular extent $\Theta^*\sim0.6^\circ-4^\circ$.

\end{document}